# Angular momentum bounds in particle systems

Daniel Pfenniger

*University of Geneva, Geneva Observatory, ch. des Maillettes 51, 1290 Versoix, Switzerland*

`daniel.pfenniger@unige.ch`

November 25$^{\text{th}}$, 2019

**Abstract**

Four expressions involving sums of position and velocity coordinates bounding the total angular momentum of particle systems, and by extension of any continuous or discontinuous material systems, are derived which are tighter for any particle configuration than similar inequalities derived by Sundman [16], Saari [14], and Scheeres [15]. Eight distinct inequalities can thus be ordered according to their tightness to angular momentum.

## 1  Introduction

As well known the *action* quantity $A$ is the fundamental scalar quantity of mechanical systems. The differential of $A$ is the difference between kinetic energy $T$ and potential energy $U$ times the time interval $dt$, $dA = (T - U)\, dt$, Any trajectory minimizes the action $A$, the Maupertuis' principle of least action. Quantum mechanics becomes relevant when $A$ is of the order or less than the quantum of action, Planck's constant $h$.

With the same unit as action but being a bivector (an oriented surface), the angular momentum quantity, $\vec{L} = m\vec{x} \wedge \vec{v}$, where $\vec{x}$ is a position, $\vec{v}$ a velocity and $m$ a mass, is the next fundamental quantity related to rotation in the continuous and isotropic space-time. Its magnitude is the spin of the system, also quantized when of order of $h$. In particular $\vec{L}$ does not depend on the specific interactions between particles, that is from $U$. Any property of angular momentum is therefore susceptible to correspond to basic constraints of interest for a wide range of problems.

Sundman's inequality [16] extended to the full $N$-body problem [6, 8] states that the magnitude of the total angular momentum $L = |\vec{L}|$ of a system of any $N$ mass particles is bounded by its total kinetic energy $T$, and its polar moment of inertia $I$ and time-derivative $\dot{I}$:

$$L^2 + \tfrac{1}{4}\, \dot{I}^2 \leq 2T\, I\,. \tag{1}$$

The equality occurs when all the particles are confined to a plane perpendicular to $\vec{L}$ [14]. Notations are explicitly described in the next section.

In many works, Sundman's inequality is applied to specific gravitational systems, such as the three-body problem. Sundman's original paper was specific to the gravitational three-body problem.

A related constraint, but less general, was stated by Poincaré [12] about the angular velocity of rotating fluids, essentially expressing the fact that for compensating outward directed centrifugal force an isolated body in equilibrium must have everywhere an inward directed gravitational force, limiting the mean angular speed. For a gravitating isolated body of mean density $\bar{\rho}$, the angular rotation frequency $\Omega$ is bounded by

$$\Omega^2 < 2\pi G \bar{\rho}. \tag{2}$$

The Poincaré inequality was improved by various authors, including the related bounds on angular momentum (e.g. [13, 9, 10]), since $L$ is proportional to $\Omega$ times the axial inertia tensor (see Eq.(15)). All these bounds are specific to some classes of gravitational systems, as witnessed by the occurence of the gravitational constant $G$.

Here we *avoid* to relate $T$ to the interaction part of the system for keeping the generality of the discussion at a broader level. Indeed, since angular momentum is a pure mass and phase space quantity, one should be able to express simple bounds with only mass-weighted moments of position and velocity differences. By keeping generality large, the results here can thus be applied to all material systems, as well in astrophysics as in terrestrial systems, to discontinuous mass distributions (such as fractal sets) or to smooth hydrodynamical systems by taking the limit $N \to \infty$ and assuming a differentiable distribution.

Sundman's inequality has been generalized to higher dimensional spaces [3], but this requires to generalize properly angular momentum in these higher dimensional spaces. Here, the discussion is kept in the frame of classical 3D Euclidian space. The inequalities discussed here can be applied also to systems not conserving $\vec{L}$, such as subsets of a larger set of interacting particles. For systems for which $\vec{L}$ is an integral of motion, these inequalities provide lower bounds in time on the combined values of the specific moments of positions and velocities; otherwise, they are valid at each respective time in an evolving system.

The fundamental reason for this generality are purely algebraic, as they follow from properties of specific antisymmetric sums of sets of real vectors. While studying this problem, more general identities involving double sums of weighted terms extending the classical Lagrange and Binet-Cauchy identities were found, which, although of larger generality than necessary for this article, are reported in Appendix II.

Saari [14, and references therein] succeeded to improve Sundman's inequality (1), by splitting the kinetic energy into two parts, $T = T_t + T_r$, where $T_t$ and $T_r$ are, respectively, the tangential and radial components of the kinetic energy with respect to the center of mass, yielding two inequations,

$$L^2 \leq 2T_t I, \tag{3}$$
$$\tfrac{1}{4}\dot{I}^2 \leq 2T_r I. \tag{4}$$

Summing Saari's pair of inequalities (3-4) side by side yields Sundman's inequality (1). The physical meaning of Saari's inequalities is more specific than Sundman's inequality and rather intuitive: For a given mass distribution (fixing $I$), Saari's inequalities (3-4) tell us that the total angular momentum is bounded by the *tangential* kinetic energy part, and the moment of inertia variation is bounded by the *radial* kinetic energy part, while Sundman's inequality (1) tells us only that the combined angular momentum and moment of inertia variation are bounded by the total kinetic energy content.

On the other hand, Scheeres [15] discovered another similar inequality, tighter than Sundman's inequality, yet distinct from Saari's inequality,

$$L^2 \leq 2T\mathcal{I}_L, \tag{5}$$

where $\mathcal{I}_L$ is the scalar formed by projecting the axial inertia tensor $\overset{\leftrightarrow}{\mathcal{I}}$ on the unit vector $\vec{u}_L$ aligned with $\vec{L}$ (i.e., $\vec{u}_L \equiv \vec{L}/|\vec{L}|$):

$$\mathcal{I}_L \equiv \vec{u}_L^{\mathrm{T}} \cdot \overset{\leftrightarrow}{\mathcal{I}} \cdot \vec{u}_L. \tag{6}$$

By a rotation of coordinates aligning one axis with $\vec{L}$, only the part of inertia with respect to that axis turns out to be important for bounding $L^2$; the rest occurring in Sundman's inequality is superfluous.

Finally, we point out that $L^2$ is a symmetric expression with respect to swapping the positions and velocity coordinates, as well as the right-hand side terms in Sundman's inequality. But Saari's and Scheeres' inequalities in right-hand sides break this symmetry: Necessarily alternative, eventually tighter expressions must exist. The objective of this paper is to explicitly define some of them.

## 2 Notations

In this section the adopted notations are described. We consider a system of $N > 1$ particles with a positive mass $m_i$, a position $\vec{x}_i$ and a velocity $\vec{v}_i \equiv \dot{\vec{x}}_i$, $i = 1 \ldots N$, moving in 3D Euclidian space. When not indicated, the sums run over all the $N$ particles. The total mass is $M \equiv \sum_i m_i$. The standard scalar and vector products are noted $\cdot$ and $\wedge$, respectively.

### 2.1 Polar moment of inertia

The polar moment of inertia $I$ commonly found in the astrophysical literature [e.g., 2, 1] is a scalar characterizing the average mass extension in ordinary space. It can be expressed in an intrinsic way, independent of the choice of origin and orientation of the coordinate system [e.g., 7, 14] by using only the relative distances between the particles. This definition is valid in arbitrarily accelerated reference frames. Note $\vec{x}_{ij} \equiv \vec{x}_i - \vec{x}_j$,

$$I \equiv \frac{1}{M} \sum_{i=1}^{N-1} \sum_{j=i+1}^{N} m_i m_j \, (\vec{x}_i - \vec{x}_j)^2 = \frac{1}{2M} \sum_{i,j} m_i m_j \, \vec{x}_{ij}^2. \tag{7}$$

By choosing $a_i \equiv \sqrt{m_i} x_i$ and $b_i \equiv \sqrt{m_i}$, we can recast the double sum over $N$ by a simple sum over $N$, using Lagrange's identity (Appendix II), which spares much computations for large $N$,

$$I = \sum_i m_i \, \vec{x}_i^2 - M\vec{X}^2, \tag{8}$$

where $\vec{X} \equiv M^{-1} \sum_i m_i \vec{x}_i$ is the system center of mass. Translating $\vec{X}$ to the origin simplifies subsequent expressions, with the advantage of reducing round-off errors in numerical computations.

Saari defines $I$ as half the commonly adopted form found in the astrophysical literature. This has the notational advantage of being similarly defined as the kinetic energy $T$, but is no longer coherent with the normalization that a unit mass at unit distance should provide a unit inertia tensor value.

The kinetic energy $T$ is actually the equivalent of half a polar moment of inertia in velocity space. As much symmetry between the position and velocity coordinates exists in the considered sums, it is useful to consider $2T$ as the analogous of $I$ in velocity space.

Time-differentiating $I$ with respect to time gives,

$$\tfrac{1}{2}\dot{I} = \sum_i m_i \, \vec{x}_i \cdot \vec{v}_i - M\vec{X} \cdot \vec{V}, \tag{9}$$

where $\vec{V} \equiv M^{-1} \sum_i m_i \vec{v}_i$ is the velocity centroid.

## 2.2 Polar moment of inertia tensor

The components $I_{kl}$, where $k$ and $l$ index the coordinates $x - y - z$,

$$\begin{aligned} I_{kl} &\equiv \frac{1}{2M} \sum_{i,j} m_i m_j \left(x_{ki} - x_{kj}\right)\left(x_{li} - x_{lj}\right) \\ &= \frac{1}{2M} \sum_{i,j} m_i m_j \, x_{kij} x_{lij} = \sum_i m_i \, x_{ki} x_{li} - M X_k X_l. \end{aligned} \tag{10}$$

form a $3 \times 3$ symmetric tensor[1] called the polar moment of inertia tensor

$$\overleftrightarrow{I} \equiv \begin{pmatrix} I_{xx} & I_{xy} & I_{zx} \\ I_{xy} & I_{yy} & I_{yz} \\ I_{zx} & I_{yz} & I_{zz} \end{pmatrix}. \tag{11}$$

The trace of this symmetric tensor is the polar moment of inertia $I$ and is invariant by a rotation/translation of the coordinate system.

Differentiating the components $I_{kl}$ gives similarly,

$$\dot{I}_{kl} = \sum_i m_i \left[x_{ki} v_{li} + x_{li} v_{ki}\right] - M \left[X_k V_l + V_k X_l\right]. \tag{12}$$

Time-differentiating this tensor again leads to the tensor Lagrange-Jacobi identity and the tensor virial theorem, but they are not required here.

---

[1] The double sum has been eliminated using Lagrange's identity.

## 2.3 Axial moment of inertia tensor and angular momentum

The axial moment of inertia tensor used in rigid body mechanics [e.g., 5, Chap. 5], often confusingly also noted $I$, expresses the inertia for rotating a system about some axis. Here, we use the symbol $\overset{\leftrightarrow}{\mathcal{I}}$ to distinguish it from $\overset{\leftrightarrow}{I}$. It reads,

$$\begin{aligned}
\overset{\leftrightarrow}{\mathcal{I}} &\equiv \begin{pmatrix} \mathcal{I}_{xx} & \mathcal{I}_{xy} & \mathcal{I}_{zx} \\ \mathcal{I}_{xy} & \mathcal{I}_{yy} & \mathcal{I}_{yz} \\ \mathcal{I}_{zx} & \mathcal{I}_{yz} & \mathcal{I}_{zz} \end{pmatrix} \equiv \begin{pmatrix} I & 0 & 0 \\ 0 & I & 0 \\ 0 & 0 & I \end{pmatrix} - \overset{\leftrightarrow}{I} \\
&= \begin{pmatrix} I_{yy}+I_{zz} & -I_{xy} & -I_{zx} \\ -I_{xy} & I_{zz}+I_{xx} & -I_{yz} \\ -I_{zx} & -I_{yz} & I_{xx}+I_{yy} \end{pmatrix}.
\end{aligned} \quad (13)$$

The trace of $\overset{\leftrightarrow}{\mathcal{I}}$ amounts to $2I$.

As for $I$, angular momentum can be defined independently of the reference frame by using relative coordinates and velocities and simplified by using Lagrange's identity:

$$\vec{L} \equiv \frac{1}{2M} \sum_{i,j} m_i m_j \, \vec{x}_{ij} \wedge \vec{v}_{ij} = \sum_i m_i \, \vec{x}_i \wedge \vec{v}_i - M\vec{X} \wedge \vec{V}. \quad (14)$$

The system angular momentum and the axial inertia tensor allow to define a system instantaneous angular velocity vector $\vec{\Omega}$ by

$$\vec{L} = \overset{\leftrightarrow}{\mathcal{I}} \cdot \vec{\Omega}. \quad (15)$$

The explicit expressions for calculating the inverse or pseudo-inverse of $\overset{\leftrightarrow}{\mathcal{I}}$ are given in Appendix III and [11].

## 2.4 Kinetic energy tensor

Exactly the same formalism used for the position components can be used for the velocity components by swapping the positions and velocities. However, the kinetic energy definition demands a factor $1/2$, which as far as possible goes as a prefactor 2 of the kinetic energy terms in the next expressions for preserving the position–velocity symmetry,

$$\begin{aligned}
2T_{kl} &\equiv \frac{1}{2M} \sum_{i,j} m_i m_j \left( v_{ki} - v_{kj} \right) \left( v_{li} - v_{lj} \right) \\
&= \frac{1}{2M} \sum_{i,j} m_i m_j \, v_{kij} \, v_{lij} = \sum_i m_i \, v_{ki} v_{li} - MV_k V_l.
\end{aligned} \quad (16)$$

where $\vec{v}_{ij} \equiv \vec{v}_i - \vec{v}_j$. The trace of $T_{kl}$ is the kinetic energy

$$2T \equiv \frac{1}{2M} \sum_{i,j} m_i m_j \, \vec{v}_{ij}^{\,2} = \sum_i m_i \, \vec{v}_i^{\,2} - M\vec{V}^2. \quad (17)$$

### 2.4.1 Kinetic energy split in two components

Each particle velocity vector can be decomposed as a sum of two orthogonal components, one purely radial from the mass center, and one purely rotational about the origin. Thus, the total kinetic energy can also be split in two respective radial and rotational components.

The radial part $T_r$ is defined by

$$2T_r \equiv \sum_i m_i v_{ri}^2, \qquad (18)$$

where $v_{ri} \equiv (\vec{v}_i \cdot \vec{x}_i)/|\vec{x}_i|$ is the signed radial velocity for particle $i$.

The rotational kinetic energy component $T_t$ is the difference between $T$ and $T_r$,

$$2T_t \equiv 2T - 2T_r = \sum_i m_i \frac{\vec{l}_i^{\,2}}{|\vec{x}_i|^2} = \sum_i m_i \vec{v}_{ti}^{\,2}, \qquad (19)$$

obtained by using the vector relation $|\vec{x} \wedge \vec{v}|^2 = \vec{x}^2 \vec{v}^2 - (\vec{x} \cdot \vec{v})^2$ and where $\vec{l}_i \equiv \vec{x}_i \wedge \vec{v}_i$ is the specific angular momentum, and $\vec{v}_{ti} \equiv \vec{l}_i/|\vec{x}_i|$ the tangential velocity.

### 2.4.2 Kinetic energy split in three components

Another distinct way to split kinetic energy of $N$ particles was proposed by [14] into three orthogonal components $\vec{w}_r$, $\vec{w}_t$, and $\vec{w}_c$ . Each velocity vector is split into:

1. A radial component proportional to the relative average rate of expansion of the system, linear in the position radius,

$$\vec{w}_{ri} \equiv \frac{\dot{I}}{2I}\vec{x}_i. \qquad (20)$$

   The associated kinetic energy can be expressed by the polar moment of inertia $I$ and its time derivative,

$$\sum_i m_i \vec{w}_{ri}^{\,2} = \frac{\dot{I}^2}{4I}. \qquad (21)$$

2. A rotational component proportional to the instantaneous average angular velocity $\vec{\Omega}$ and linear in the position radius,

$$\vec{w}_{ti} \equiv \vec{\Omega} \wedge \vec{x}_i, \qquad (22)$$

   where $\vec{\Omega}$ is determined by the system angular momentum $\vec{L}$ and the inverse (or pseudo-inverse) axial moment of inertia tensor

$$\vec{\Omega} = \overleftrightarrow{\mathcal{I}}^{\dagger} \cdot \vec{L}. \qquad (23)$$

The associated kinetic energy depends only on the total angular momentum and the axial inertia tensor,

$$\sum_i m_i \vec{w}_{ti}^{\,2} = \vec{L}^{\mathrm{T}} \cdot \overset{\leftrightarrow}{\mathcal{I}}{}^{\dagger} \cdot \vec{L} = \vec{\Omega}^{\mathrm{T}} \cdot \overset{\leftrightarrow}{\mathcal{I}} \cdot \vec{\Omega}. \tag{24}$$

3. A "configurational" component contains the leftover part of velocity which is not included in the two previous ones,

$$\vec{w}_{ci} \equiv \vec{v}_i - \vec{w}_r - \vec{w}_t. \tag{25}$$

These three components are shown to be "orthogonal" for the *system scalar product* [14]:

$$\langle \vec{A}, \vec{B} \rangle \equiv \sum_i m_i\, \vec{a}_i \cdot \vec{b}_i, \tag{26}$$

where $\vec{A} \equiv (\vec{a}_1, \vec{a}_2, \ldots \vec{a}_N)$, $\vec{B} \equiv \left(\vec{b}_1, \vec{b}_2, \ldots \vec{b}_N\right)$. Thus

$$\langle \vec{\vec{w}}_r, \vec{w}_t \rangle = \langle \vec{\vec{w}}_t, \vec{w}_c \rangle = \langle \vec{\vec{w}}_c, \vec{w}_r \rangle = 0 \tag{27}$$

and

$$2T = \langle \vec{\vec{w}}_r, \vec{w}_r \rangle + \langle \vec{\vec{w}}_t, \vec{w}_t \rangle + \langle \vec{\vec{w}}_c, \vec{w}_c \rangle. \tag{28}$$

Expressing the total kinetic energy with the above kinetic energy components, an *equality* links it with the total angular momentum and the polar moment of inertia and the axial inertia tensors [e.g., 4],

$$2T \;=\; \frac{\dot{I}^2}{4I} + \vec{L}^{\mathrm{T}} \cdot \overset{\leftrightarrow}{\mathcal{I}}{}^{\dagger} \cdot \vec{L} + \sum_i m_i \vec{w}_{ci}^{\,2}. \tag{29}$$

This equality is interesting as it decomposes kinetic energy in three natural components on the right-hand side, respectively: the energy linked with the average rate of bulk radial motion, the energy associated with the average bulk rotation and the leftover energy related in many cases to disordered, thermal-like motion.

## 3 Tightening Sundman's inequality

Now let us derive six tighter inequalities involving the angular momentum components and the components of the diagonal of the inertia and kinetic energy tensors.

We start from the fast derivation of Sundman's inequality [6, p. 28–31] achieved by using the algebraic general inequality given in Appendix I, valid for *any* pair of sets of three-vectors $\{\vec{a}_i\}$ and $\{\vec{b}_i\}$, $i = 1 \ldots N$,

$$\left(\sum_i \vec{a}_i \wedge \vec{b}_i\right)^2 + \left(\sum_i \vec{a}_i \cdot \vec{b}_i\right)^2 \leq \sum_i \vec{a}_i^{\,2} \sum_j \vec{b}_j^{\,2}. \tag{30}$$

By setting $\vec{a}_i \equiv \sqrt{m_i}\vec{x}_i$ and $\vec{b}_i \equiv \sqrt{m_i}\vec{v}_i$, Sundman's inequality (1) is immediately established.

## 3.1 Fixing one coordinate to zero

Now a key argument used here is that since this inequality is valid for *any* $\{\vec{a}_i\}$ and $\{\vec{b}_i\}$, we have the freedom to choose to set, say, the $z$-component of each $\{\vec{a}_i\}$ and $\{\vec{b}_i\}$ to zero, instead of the effective $z$-component of the particles. This gives,

$$\left[\sum_i \left(a_{xi}b_{yi} - a_{yi}b_{xi}\right)\right]^2 + \left[\sum_i \left(a_{xi}b_{xi} + a_{yi}b_{yi}\right)\right]^2 \leq \sum_i \left(a_{xi}^2 + a_{yi}^2\right) \sum_j \left(b_{xj}^2 + b_{yj}^2\right) \tag{31}$$

This can be seen as a particular form of (30) valid for any pair of 2-vector sets $\{\vec{a}_i\}$, $\{\vec{b}_i\}$.

Substituting now the physical values gives a constraint on the $L_z$ component of the angular momentum alone. Likewise, cycling over the three axes gives a constraint for each component of angular momentum:

$$\begin{aligned}
L_x^2 + \tfrac{1}{4}\left(\dot{I}_{yy} + \dot{I}_{zz}\right)^2 &\leq 2\left(T_{yy} + T_{zz}\right)\left(I_{yy} + I_{zz}\right), \\
L_y^2 + \tfrac{1}{4}\left(\dot{I}_{zz} + \dot{I}_{xx}\right)^2 &\leq 2\left(T_{zz} + T_{xx}\right)\left(I_{zz} + I_{xx}\right), \\
L_z^2 + \tfrac{1}{4}\left(\dot{I}_{xx} + \dot{I}_{yy}\right)^2 &\leq 2\left(T_{xx} + T_{yy}\right)\left(I_{xx} + I_{yy}\right).
\end{aligned} \tag{32}$$

These three inequalities give constraints on the respective components of $\vec{L}$. Rotating the system such that only one component, for example $L_z$, is nonzero gives obviously tighter inequations (32) than the original scalar inequality (1), since the irrelevant particle coordinates, for a given angular momentum component, are absent. Retrospectively, it seems clear that if the minimizing quantity, for example $L_z^2$, does not depend on the $\{z_i, v_{zi}\}$ coordinates, the inequality must hold mathematically also when these quantities are set to any particular values, and in particular to those producing the tightest inequation, $z_i = 0, v_{zi} = 0$. The coordinate rotation to align $\vec{L}$ with the $z$-axis serves only to make the algebraic expressions simpler.

Summing the three inequalities in (32) to build a single inequality on $L^2$ provides a poorer inequality than Sundman's. For example, inspection reveals it is always poorer when the system is in average stationary ($\dot{I} = 0$).

The improvement on the relative bound given by (32) over (1) for a spherical isotropic and steady system ($I_{xx} = I_{yy} = I_{zz}$, $T_{xx} = T_{yy} = T_{zz}$, $\dot{I}_{xx} = \dot{I}_{yy} = \dot{I}_{zz} = 0$) amounts to 1/3. The improvement is higher for systems elongated along the rotation axis, like prolate spheroids, and lower for systems flattened along the rotation axis, like oblate spheroids. For purely planar distributions, the relevant component of (32) is equivalent to (1).

## 3.2 Fixing two coordinates to zero

Choosing now to set two coordinates to zero of each of the $\{\vec{a}_i\}$ and $\{\vec{b}_i\}$, say the $y$ and $z$ components, gives actually a Cauchy-Schwarz inequality,

$$\left(\sum_i a_{xi} b_{xi}\right)^2 \leq \sum_i a_{xi}^2 \sum_j b_{xj}^2. \tag{33}$$

Cycling over the $x-y-z$ coordinates, we obtain three inequalities constraining the rate of change in magnitude of each diagonal component of $I$,

$$\tfrac{1}{4}\dot{I}_{xx}^2 \leq 2T_{xx}I_{xx}, \qquad \tfrac{1}{4}\dot{I}_{yy}^2 \leq 2T_{yy}I_{yy}, \qquad \tfrac{1}{4}\dot{I}_{zz}^2 \leq 2T_{zz}I_{zz}. \tag{34}$$

Again these inequalities are valid for any $N$-body system, not only to systems confined on a straight line. These inequalities are specialized forms of Saari's second inequality, and equivalent for linear systems aligned along one axis.

An application of these inequalities is to give an instantaneous lower bound on the timescale ($\tau_k \equiv I_{kk}/\dot{I}_{kk}$) for significant change in each respective direction based only on the knowledge of the present size and kinetic energy content:

$$\tau_x > \sqrt{\frac{I_{xx}}{8T_{xx}}}, \qquad \tau_y > \sqrt{\frac{I_{yy}}{8T_{yy}}}, \qquad \tau_z > \sqrt{\frac{I_{zz}}{8T_{zz}}}. \tag{35}$$

## 4 Tightening Saari's inequalities

Now we repeat the method used for Sundman's inequality on Saari's inequalities by splitting them in coordinate components.

Looking at the demonstrations of these inequalities is instructive. The first inequality Equ.(3) demonstration goes as follows (somewhat shorter than the demonstration given by [14, p. 62]):

$$L^2 = \left(\sum_i m_i \vec{x}_i \wedge \vec{v}_i\right)^2 = \left(\sum_i \left[\sqrt{m_i}\frac{\vec{x}_i \wedge \vec{v}_i}{|\vec{x}_i|}\right][\sqrt{m_i}\,|\vec{x}_i|]\right)^2. \tag{36}$$

Using Cauchy-Schwarz inequality on the terms inside brackets, we obtain,

$$L^2 \leq \sum_i m_i \frac{|\vec{x}_i \wedge \vec{v}_i|^2}{|\vec{x}_i|^2} \sum_i m_i |\vec{x}_i|^2 = 2T_t\, I. \tag{37}$$

The demonstration of the second inequality Equ.(4) is even shorter,

$$\tfrac{1}{2}\dot{I} = \sum_i m_i \vec{x}_i \cdot \vec{v}_i = \sum_i m_i |\vec{x}_i| v_{ri} = \sum_i [\sqrt{m_i} v_{ri}][\sqrt{m_i}\,|\vec{x}_i|]. \tag{38}$$

Squaring and using Cauchy-Schwarz inequality on the terms inside bracket give the result,

$$\tfrac{1}{4}\dot{I}^2 \leq 2T_r I. \tag{39}$$

## 4.1 Fixing one coordinate to zero

Here, we use a similar procedure as for Sundman's inequality for splitting Saari's inequalities into vector components, giving still tighter constraints on the magnitude of each component of $\vec{L}$.

Considering just one component, say $L_z$, the result will as well apply to the other components by cyclic permutation of $x - y - z$. Note the radius in the $x - y$ space by $R_z \equiv \sqrt{x^2 + y^2}$, and the, respectively, signed radial and tangential velocities with respect to the $z$-axis by

$$v_{R_z} \equiv \frac{xv_x + yv_y}{R_z}, \quad v_{t_z} \equiv \frac{xv_y - yv_x}{R_z} = \frac{l_z}{R_z}. \tag{40}$$

We define also $T_{R_z}$ and $T_{t_z}$ as, respectively, the radial and tangential kinetic energies with respect to the $z$-axis,

$$2T_{R_z} \equiv \sum_i m_i v_{R_z i}^2, \qquad 2T_{t_z} \equiv \sum_i m_i v_{t_z i}^2. \tag{41}$$

We have

$$T_{xx} + T_{yy} = T_{R_z} + T_{t_z}. \tag{42}$$

Now we can use a similar decomposition for just one vector component as done above for the angular momentum vector length,

$$L_z = \sum_i m_i \left( x_i v_{y_i} - y_i v_{x i} \right) = \sum_i \left[ \sqrt{m_i} \, v_{t_z i} \right] \left[ \sqrt{m_i} \, R_{z i} \right]. \tag{43}$$

Squaring and using again the Cauchy-Schwarz inequality over the terms inside brackets, we obtain,

$$L_z^2 \leq \sum_i m_i v_{t_z i}^2 \sum_i m_i R_{z i}^2 = 2T_{t_z} \left( I_{xx} + I_{yy} \right). \tag{44}$$

Cycling over the axes $x - y - z$, the improved component inequalities read,

$$\begin{array}{rcl}
L_x^2 & \leq & 2T_{t_x} \left( I_{yy} + I_{zz} \right) = 2T_{t_x} \mathcal{I}_{xx}, \\
L_y^2 & \leq & 2T_{t_y} \left( I_{zz} + I_{xx} \right) = 2T_{t_y} \mathcal{I}_{yy}, \\
L_z^2 & \leq & 2T_{t_z} \left( I_{xx} + I_{yy} \right) = 2T_{t_z} \mathcal{I}_{zz}.
\end{array} \tag{45}$$

For the same reason as for Sundman's inequality, the respective component inequalities are tighter than the scalar inequality. This is seen by rotating the coordinates such that one axis is parallel to $\vec{L}$. However summing these three inequalities does not yield a better inequality than Saari's. Inequalities (45) are the tightest obtained yet.

The meaning of (45) is that the angular momentum magnitude around each principal axis is bounded by the product of the axial moment of inertia and tangential kinetic energy related to the respective axis.

The corresponding form of Saari's second inequality (4) is obtained by setting to zero all the matching position–velocity coordinates, say $\{z_i = 0, v_{zi} = 0\}$:

$$\tfrac{1}{2}\dot{\mathcal{I}}_{zz} = \sum_i m_i \left(x_i v_{xi} + y_i v_{y_i}\right) = \sum_i \left[\sqrt{m_i} v_{R_z i}\right] \left[\sqrt{m_i} R_{zi}\right]. \tag{46}$$

Squaring and using the Cauchy-Schwarz inequality, splitting the last sum with the bracketed terms and cycling over the coordinates, yield the component version of Saari's second inequality (4),

$$\tfrac{1}{4}\dot{\mathcal{I}}_{xx}^2 \leq 2T_{R_x}\mathcal{I}_{xx}, \quad \tfrac{1}{4}\dot{\mathcal{I}}_{yy}^2 \leq 2T_{R_y}\mathcal{I}_{yy}, \quad \tfrac{1}{4}\dot{\mathcal{I}}_{zz}^2 \leq 2T_{R_z}\mathcal{I}_{zz}. \tag{47}$$

Summing the respective component inequalities (45) and (47) gives back the improved Sundman's component inequalities (32).

The physical meaning of (47) is that the rate of change in the axial moment of inertia is bounded by the product of the axial moment of inertia with the axis-related radial kinetic energy.

## 4.2 Fixing two coordinates to zero

If one now considers only the $x$ component of $\dot{I}$ in Equ. (38), one obtains

$$\tfrac{1}{2}\dot{I}_{xx} = \sum_i m_i x_i \cdot v_{xi} = \sum_i [\sqrt{m_i} x_i][\sqrt{m_i} v_{xi}]. \tag{48}$$

Squaring and using Cauchy-Schwarz inequality give the same result as Equ. (34).

## 5 Last improvements

If we consider the expressions of $L_x^2$, $L_y^2$ and $L_z^2$ in terms of particle coordinates, one notes that they are completely symmetric with respect to swapping the positions and velocities; the quantities called positions could be called velocities and vice versa without changing the algebraic results.

But on the other hand, in inequations (45) the right-hand sides are *not* symmetric with respect to such a position–velocity swap. Therefore one can in principle obtain distinct inequalities where the right-hand side contains expressions where the position and velocity coordinates are swapped too.

For example, the position–velocity swap in the $L_z^2$ inequality (44) reads:

$$\begin{aligned} L_z^2 &\leq \sum_i m_i \left(v_{xi}^2 + v_{yi}^2\right) \sum_i m_i \left[\frac{l_{zi}^2}{v_{xi}^2 + v_{yi}^2}\right] \\ &= 2\left(T_{xx} + T_{yy}\right) \sum_i m_i \left[\frac{l_{zi}^2}{v_{xi}^2 + v_{yi}^2}\right] \\ &= 2\left(T_{R_z} + T_{t_z}\right) \sum_i m_i \frac{R_{zi}^2}{1 + \frac{v_{R_z i}^2}{v_{t_z i}^2}}. \end{aligned} \tag{49}$$

By the symmetry principle, since these new expressions are distinct, they generally entail different values, while anyone cannot be systematically smaller than the other for the same symmetry reason. This means that for any particular case, one of the two possibilities has a tighter bound than the other, and the smallest one can be chosen.

So the last improvement in a particular case is to take both expressions and keep the smallest of the two possibilities. Calling the first possibility $p_1$ and the second $p_2$, a formal expression can be written as

$$L_z^2 \leq \min(p_1, p_2) = \tfrac{1}{2}\left(p_1 + p_2 - |p_1 - p_2|\right). \tag{50}$$

This bound is in principle the best bound of this paper on the $\vec{L}$-component magnitudes expressed with only the respective weighted moments of positions and velocities, although, in view of the more complicated expressions, less practical than the other ones.

# 6 Summary of $|\vec{L}|$ bounds

In summary, when the angular momentum $\vec{L}$ is aligned with the $z$-axis, the square rooted bounds discussed above on $L = |L_z| = |\vec{L}|$ are:

1. The Sundman bound Equ. (1), named Su:

    $$L \leq \sqrt{2T\,I}.$$

2. The Saari bound Equ. (3), named Sa:

    $$L \leq \sqrt{2T_t\,I}.$$

    Since $T_t \leq T$, Sa $\leq$ Su.

3. The Scheeres bound Equ. (5), named Sc:

    $$L \leq \sqrt{2T\,\mathcal{I}_{zz}}.$$

    Since $\mathcal{I}_{zz} \leq I$, Sc $\leq$ Su, but no particular order between Sa and Sc can be stated.

4. The minimum between Saari and Scheeres bounds, named Sac. Thus, Sac $\leq$ Sa, and Sac $\leq$ Sc.

5. The improved $z$-component bound in Equ. (32), named $P_1$:

    $$L \leq \sqrt{2\,(T_{xx} + T_{yy})(I_{xx} + I_{yy})} = \sqrt{2\,(T_{R_z} + T_{t_z})\,\mathcal{I}_{zz}}.$$

    Since $I_{xx} + I_{yy} \leq I$, and $T_{xx} + T_{yy} \leq T$, $P_1 \leq$ Sa and $P_1 \leq$ Sc, $P_1 \leq$ Sac.

6. The improved $z$-component bound in Equ. (45), named P$_2$:

$$L \leq \sqrt{2T_{t_z}\mathcal{I}_{zz}}.$$

Since $T_{t_z} \leq T_{t_z} + T_{R_z}$, P$_2 \leq$ P$_1$.

7. The position–velocity swapped version of the previous bound in Equ. (49), named P$_3$:

$$L \leq \sqrt{2\left(T_{xx} + T_{yy}\right)\sum_i m_i R_{zi}^2/(1 + v_{R_z i}^2/v_{t_z i}^2)}.$$

By symmetry, P$_2$ and P$_3$ cannot be systematically smaller or larger than the other.

8. The minimum between the last two bounds, named P$_{23}$. By construction, P$_{23}$ is the smallest bound.

Gathering these results, the bounds are ordered as follows:

$$L = |L_z| \leq \text{P}_{23} \leq \begin{matrix} \text{P}_2 \\ \text{P}_3 \end{matrix} \leq \text{P}_1 \leq \text{Sac} \leq \begin{matrix} \text{Sa} \\ \text{Sc} \end{matrix} \leq \text{Su}. \tag{51}$$

Finally, replacing in P$_3$ $(T_{xx} + T_{yy})$ by $T_{t_z}$ does not provide a tighter inequality, as specific examples checked numerically are either smaller or larger than $L$.

# 7 Numerical check

A simple numerical check is set up to yield quantitative differences that may occur on the various angular momentum bounds discussed above. We draw randomly $N$-body configurations with $N$ rather small in the range $(4 \leq N \leq 10)$ to well explore the possible conditions. Indeed, when $N$ is large the derived distributions reflect more the particularly chosen random distributions than the range of possibilities. Keeping $N$ small allows to better probe the effect of large statistical fluctuations.

A sample of $10^8$ such $N$-body configurations is drawn, where $N$ is uniformly random between 4 and 10 included. The masses are drawn from a uniform distribution over $[0 - 1]$ and normalized to a unit total mass. The positions and velocities are drawn from an ellipsoidal normal distribution with the same axis ratios $2/1/0.5$ in position and velocity. Each drawn $N$-body configuration is shifted such that the centers of mass and velocity are at the origin. Then, a uniform rotation is applied to velocities, $\vec{v}_i' = \vec{v}_i + \vec{\delta\Omega} \wedge \vec{x}_i$, such that the final total angular momentum $\vec{L}' = \vec{L} + \mathcal{I} \cdot \vec{\delta\Omega}$ is unity and oriented along the $z$-axis. This keeps the generality of the check since a nonzero angular momentum can always be normalized by a change of time unit. Doing so also excludes cases with zero angular momentum. The ratio of rotational to random kinetic energy is determined by the initial amplitude of the random velocities.

The bounds enumerated in Sect. 6 are calculated. The largest of two bounds is set at the numerator, the smallest to the denominator, so that the log of the ratio is always positive. Then, all possible nonredundant histograms of relevant ratios in logarithmic scale are inspected, showing that the bound order discussed in Sect. 6 holds for the sample. The bound pairs Sa–Sc, and $P_2$–$P_3$ (not shown) can display any positive ratio, so their log can be positive or negative. The most relevant histograms are displayed in Fig. 1.

The first percentile and the median of the ratios to the angular momentum are indicated at the top of the frames. For example, for half of the $10^8$ configurations, the Sundman bound Su is less than 3.96 times the angular momentum magnitude, while this median ratio drops to 1.80 for the tightest bound $P_{23}$. For the three tightest bounds $P_2$, $P_3$, $P_{23}$ one observes that their distribution is concave before their maximum, contrary to the other bounds, which improves drastically their low percentiles.

## 8   Conclusions

Eight bounds on the total angular momentum of a system that can be inferred from various global mass weighted moments of the positions and velocities have been discussed and presented. Four new and tighter bounds with respect to previous known bounds have been found ($P_1$,$P_2$,$P_3$ and $P_{23}$), which yield insight on the average quantities that constrain the total angular momentum at best. In a particular sample of randomly drawn $10^8$ $N$-body systems, the improvement with respect to the known previous bounds is substantial: while the median ratio of the previous bounds to angular momentum was 3.56 or more, a new bound ($P_1$) improves this median ratio to 3.36, the new bounds $P_2$, $P_3$ improves it to 1.98 and finally the bound $P_{23}$ to 1.80.

## Appendix I: Hill's inequations

In Hill's book [6, p. 29–30] is proved an important double inequation that we display here in more current notations. For any pair of real three-vector sets $\{\vec{a}_i\}$ and $\{\vec{b}_i\}$, $i = 1 \ldots N$, then

$$\sum_i \vec{a}_i^2 \sum_j \vec{b}_j^2 - \left(\sum_i \vec{a}_i \cdot \vec{b}_i\right)^2 \geq \left(\sum_i \left|\vec{a}_i \wedge \vec{b}_i\right|\right)^2 \qquad (52)$$

$$\geq \left(\sum_i \vec{a}_i \wedge \vec{b}_i\right)^2. \qquad (53)$$

The demonstration uses the vector relations

$$\vec{a} \cdot \vec{b} = ab\cos\theta \quad \text{and} \quad \left|\vec{a} \wedge \vec{b}\right| = ab\sin\theta \qquad (54)$$

where $\theta$ is the angle between the vectors $\vec{a}$ and $\vec{b}$, and $a$, $b$ are the lengths of $\vec{a}$ and $\vec{b}$.

Thus, subtracting the right side from the left side in Equ. (52), expanding the squared sums in double sums, and using the trigonometric identity $\cos(\theta - \eta) = \cos(\theta)\cos(\eta) + \sin(\theta)\sin(\eta)$, we obtain,

$$\sum_{i,j} a_i^2 b_j^2 - a_i b_i a_j b_j \cos\theta_i \cos\theta_j - a_i b_i a_j b_j \sin\theta_i \sin\theta_j \qquad (55)$$

$$= \sum_{i,j} a_i^2 b_j^2 - a_i b_i a_j b_j \cos(\theta_i - \theta_j). \qquad (56)$$

Each diagonal term with $i = j$ vanishes, and each pair of nondiagonal terms $i \neq j$ provides a nonnegative contribution to the sum,

$$a_i^2 b_j^2 + a_j^2 b_i^2 - 2 a_i b_j a_j b_i \cos(\theta_i - \theta_j) \geq (a_i b_j - a_j b_i)^2 \geq 0. \qquad (57)$$

The equality in Equ. (52) is reached when all the ratios $a_i/b_j$ are equal, and all the angles $\theta_i$ are equal.

Finally, Equ. (53) follows from the triangle inequality $|a| + |b| \geq |a + b|$.

Using the triangle inequality for the quadratic norm, $(|a| + |b|)^2 \geq a^2 + b^2$, one can also express another inequality

$$\sum_i \vec{a}_i^2 \sum_j \vec{b}_j^2 - \left(\sum_i \vec{a}_i \cdot \vec{b}_i\right)^2 \geq \left(\sum_i \left|\vec{a}_i \wedge \vec{b}_i\right|\right)^2 \qquad (58)$$

$$\geq \sum_i \left(\vec{a}_i \wedge \vec{b}_i\right)^2, \qquad (59)$$

which is distinct from Equ.(53), that is, in different cases either the right-hand side of Equ. (53) or that of Equ. (59) is the largest term.

## Appendix II: Weighted Binet-Cauchy identities

For $m$ pairs of real or complex $N$-dimensional vectors $a_{ki}$ and $b_{ki}$, $k \in [1, \ldots m]$, and the real or complex $N$-dimensional vector $w_i$, $i \in [1, \ldots N]$, we have the following simplifications of antisymmetric double sums.

### 1. Odd $m$

If $m$ is odd,

$$\sum_{i=1, j=1}^{N} w_i w_j \prod_{k=1}^{m} (a_{ki} b_{kj} - a_{kj} b_{ki}) = 0. \tag{60}$$

This follows from the antisymmetry of the global sum, each term occurring twice with opposite signs.

### 2. Even $m$

If $m$ is even,

$$\sum_{i=1, j=1}^{N} w_i w_j \prod_{k=1}^{m} (a_{ki} b_{kj} - a_{kj} b_{ki}) = 2 \sum_{i=1}^{N-1} \sum_{j=i+1}^{N} w_i w_j \prod_{k=1}^{m} (a_{ki} b_{kj} - a_{kj} b_{ki}). \tag{61}$$

This follows from the symmetry of the global sum, each term occurring twice with same signs.

### 3. $m = 2$

For $m = 2$, we have a weighted Binet-Cauchy identity:

$$\sum_{i=1}^{N-1} \sum_{j=i+1}^{N} w_i w_j (a_{1i} b_{1j} - a_{1j} b_{1i})(a_{2i} b_{2j} - a_{2j} b_{2i}) =$$
$$\sum_i w_i a_{1i} a_{2i} \sum_j w_j b_{1j} b_{2j} - \sum_i w_i a_{1i} b_{2i} \sum_j w_j a_{2j} b_{1j}. \tag{62}$$

If $b_{1i} = b_{2i}$, we have, dropping the index $k$ for the $b$'s:

$$\sum_{i=1}^{N-1} \sum_{j=i+1}^{N} w_i w_j (a_{1i} b_j - a_{1j} b_i)(a_{2i} b_j - a_{1j} b_i) =$$
$$\sum_i w_i a_{1i} a_{2i} \sum_j w_j b_j^2 - \sum_i w_i a_{1i} b_i \sum_j w_j a_{2j} b_j. \tag{63}$$

If $a_{1i} = \bar{a}_{2i}$ and $b_{1i} = \bar{b}_{2i}$, we have a weighted version of Lagrange's identity:

$$\sum_{i=1}^{N-1} \sum_{j=i+1}^{N} w_i w_j |a_i b_j - a_j b_i|^2 = \sum_i w_i |a_i|^2 \sum_j w_j |b_j|^2 - \left| \sum_i w_i a_i b_i \right|^2 \tag{64}$$

For real nonnegative weights $w_i$, since the left-hand side is nonnegative, a weighted form of Cauchy-Schwarz inequality follows:

$$\sum_i w_i |a_i|^2 \sum_j w_j |b_j|^2 \geq \left| \sum_i w_i a_i b_i \right|^2. \qquad (65)$$

Further, if $b_i = 1$, we have

$$\sum_{i=1}^{N-1} \sum_{j=i+1}^{N} w_i w_j |a_i - a_j|^2 = \sum_i w_i |a_i|^2 \sum_j w_j - \left| \sum_i w_i a_i \right|^2. \qquad (66)$$

For larger even $m$, it is straightforward to derive further similar equalities and inequalities.

## Appendix III

We can calculate the inverse of the axial moment of inertia, $\overset{\leftrightarrow}{\mathcal{I}}^{-1}$ (or pseudo-inverse $\overset{\leftrightarrow}{\mathcal{I}}^{\dagger}$ for degenerate cases) with the following procedure working in all configurations [11]:

1. Calculate the determinant $\left| \overset{\leftrightarrow}{\mathcal{I}} \right|$ with

$$\left| \overset{\leftrightarrow}{\mathcal{I}} \right| = \mathcal{I}_{xx} \mathcal{I}_{yy} \mathcal{I}_{zz} + 2 \mathcal{I}_{xy} \mathcal{I}_{yz} \mathcal{I}_{zx} - \left( \mathcal{I}_{xx} \mathcal{I}_{yz}^2 + \mathcal{I}_{yy} \mathcal{I}_{zx}^2 + \mathcal{I}_{zz} \mathcal{I}_{xy}^2 \right). (67)$$

2. - If $\left| \overset{\leftrightarrow}{\mathcal{I}} \right| \neq 0$, the general case, then

$$\overset{\leftrightarrow}{\mathcal{I}}^{-1} = \left| \overset{\leftrightarrow}{\mathcal{I}} \right|^{-1} \times \begin{pmatrix} \mathcal{I}_{yy}\mathcal{I}_{zz} - \mathcal{I}_{yz}^2 & \mathcal{I}_{yz}\mathcal{I}_{zx} - \mathcal{I}_{xy}\mathcal{I}_{zz} & \mathcal{I}_{xy}\mathcal{I}_{yz} - \mathcal{I}_{zx}\mathcal{I}_{yy} \\ \mathcal{I}_{yz}\mathcal{I}_{zx} - \mathcal{I}_{xy}\mathcal{I}_{zz} & \mathcal{I}_{zz}\mathcal{I}_{xx} - \mathcal{I}_{zx}^2 & \mathcal{I}_{zx}\mathcal{I}_{xy} - \mathcal{I}_{yz}\mathcal{I}_{xx} \\ \mathcal{I}_{xy}\mathcal{I}_{yz} - \mathcal{I}_{zx}\mathcal{I}_{yy} & \mathcal{I}_{zx}\mathcal{I}_{xy} - \mathcal{I}_{yz}\mathcal{I}_{xx} & \mathcal{I}_{xx}\mathcal{I}_{yy} - \mathcal{I}_{xy}^2 \end{pmatrix} \quad (68)$$

   - Else, i.e., $\left| \overset{\leftrightarrow}{\mathcal{I}} \right| = 0$, all particles are located along a line:
     (a) If $I \neq 0$, then
     $$\overset{\leftrightarrow}{\mathcal{I}}^{\dagger} = I^{-2} \overset{\leftrightarrow}{\mathcal{I}}, \qquad (69)$$
     (b) Else, i.e., $I = 0$, all the particles are at the origin,
     $$\overset{\leftrightarrow}{\mathcal{I}}^{\dagger} = 0. \qquad (70)$$

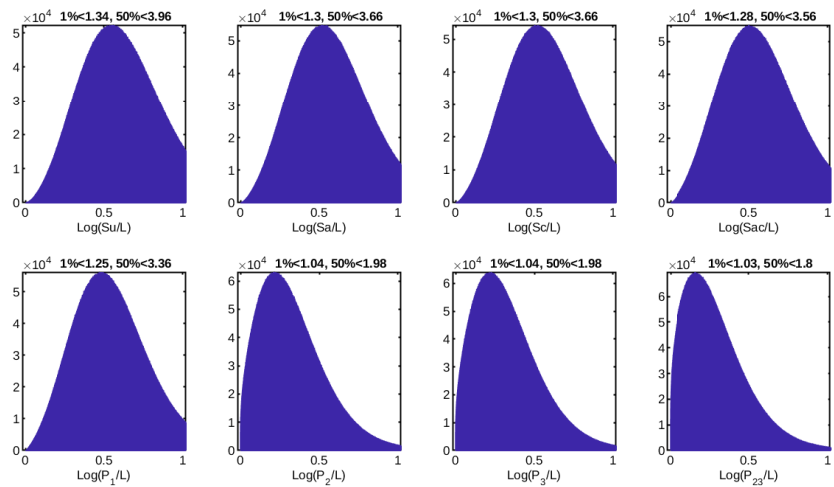

Figure 1: Histograms of bound to angular momentum ratios (the absissa is in logarithmic scale) for $10^8$ randomly chosen $N$-body configurations as decribed in the text. At the top of the frames the 1% and 50% percentiles of the ratios are indicated.

19